\documentclass[aps,pra,twocolumn,superscriptaddress,showpacs]{revtex4}

\usepackage{amsmath}
\usepackage{amssymb}
\usepackage{graphicx}
\usepackage{epsfig}
\usepackage{dcolumn}
\usepackage{multirow}
\usepackage{bbm}
\usepackage{color}
\usepackage[ps2pdf,
bookmarks=true,
bookmarksnumbered=true,
hypertexnames=false,
]{hyperref}

\newcommand{\bra}[1]{\left< #1 \right|}
\newcommand{\ket}[1]{\left| #1 \right>}
\newcommand{\sbra}[1]{\left( #1 \right|}
\newcommand{\sket}[1]{\left| #1 \right)}
\newcommand{\expect}[1]{\left< #1 \right>}
\newcommand{\braket}[2]{\left<\left.\! #1 \right|\! #2 \right>}

\newcommand{\IDM}[0]{\text{IDM}}

\newcommand{\half}[0]{\frac{1}{2}}

\newcommand{\sqrthalf}[0]{\frac{1}{\sqrt{2}}}

\newcommand{\mytitle}{Theory of Attosecond Transient Absorption Spectroscopy 
of Krypton for Overlapping Pump and Probe Pulses}

\newcommand{\rmpdfinfo}{\special{ps:: userdict /pdfmark /cleartomark load put}}

\definecolor{MyDarkGreen}{rgb}{0,0.6,0}
\definecolor{MyDarkBlue}{rgb}{0,0,0.8}
\definecolor{MyDarkRed}{rgb}{0.6,0,0.3}

\hypersetup{breaklinks=true,
colorlinks =true,
plainpages =true,
linktocpage=true,
linkcolor=MyDarkBlue,
citecolor=MyDarkGreen,
urlcolor =MyDarkRed,
pdfborder={0 0 0},
pdfauthor={Stefan Pabst},
pdftitle ={\mytitle},
pdfsubject  = {Research Article},
pdfkeywords = {Attosecond Physics, Strong-Field Physics, Transient Absorption,
               Atomic Krypton, Pump-Probe Experiment, TDCIS, Multi-Channel, 
               Many-Body Correlation Effects},
pdfcreator  = {LaTeX with hyperref package},
pdfproducer = {dvipdf}
}

\begin{document} 

\title{\mytitle}

\author{Stefan Pabst}
\affiliation{Center for Free-Electron Laser Science, DESY, Notkestrasse 85, 22607 Hamburg, Germany}
\affiliation{Department of Physics,University of Hamburg, Jungiusstrasse 9, 20355 Hamburg, Germany}

\author{Arina Sytcheva}
\affiliation{Center for Free-Electron Laser Science, DESY, Notkestrasse 85, 22607 Hamburg, Germany}

\author{Antoine Moulet}
\affiliation{Max-Planck-Institut f\"ur Quantenoptik, Hans-Kopfermann-Str. 1, D-85748 Garching, Germany}

\author{Adrian Wirth}
\affiliation{Max-Planck-Institut f\"ur Quantenoptik, Hans-Kopfermann-Str. 1, D-85748 Garching, Germany}

\author{Eleftherios Goulielmakis}
\affiliation{Max-Planck-Institut f\"ur Quantenoptik, Hans-Kopfermann-Str. 1, D-85748 Garching, Germany}

\author{Robin Santra}
\affiliation{Center for Free-Electron Laser Science, DESY, Notkestrasse 85, 22607 Hamburg, Germany}
\affiliation{Department of Physics,University of Hamburg, Jungiusstrasse 9, 20355 Hamburg, Germany}
\thanks{Corresponding author}
\email{robin.santra@cfel.de}

\date{\today}

\begin{abstract}
We present the first fully \textit{ab initio} calculations for 
attosecond transient absorption spectroscopy of atomic krypton with 
overlapping pump and probe pulses.
Within the time-dependent configuration interaction singles (TDCIS) approach, 
we describe the pump step (strong-field ionization using a near-infrared pulse) as well as the probe step 
(resonant electron excitation using an extreme-ultraviolet pulse) from first principles. 
We extent our TDCIS model and account for the spin-orbit splitting of the occupied orbitals.
We discuss the spectral features seen in a recent attosecond transient absorption 
experiment [A. Wirth {\it et al.}, Science {\bf 334}, 195 (2011)].
Our results support the concept that the transient absorption signal can be 
directly related to the instantaneous hole population even during the ionizing 
pump pulse.
Furthermore, we find strong deformations in the absorption lines when the overlap of pump and probe pulses is maximum.
These deformations can be described by relative phase shifts in the oscillating ionic dipole.
We discuss possible mechanisms contributing to these phase shifts.
Our finding suggests that the non-perturbative laser dressing of the entire $N$-electron wave function is the main contributor.
\end{abstract}
  
  
\maketitle
  
  
\section{Introduction}
\label{sec:intro}

The interaction of matter with light is a key process in physical systems
on any length scale. 
The fundamentals of matter-light interaction can be best studied in atomic
systems due to their relative simplicity.
The absorption of light promotes electrons into excited states.
If enough energy is absorbed by the system, one or more electrons can leave
the atom, i.e., ionization takes place~\cite{St-Springer-1980,Ru-RMP-1968,BrKr-RMP-2000}.
The most common types of ionization are:
single-photon and few-photon ionizations~\cite{ScYa-Science-2010,KlHu-PRL-2011,SaKe-Nature465-2010}, 
above-threshold ionization~\cite{ScKu-PRL-1993,LeMa-PRA-2002,BuBe-PRL-2008,BuMi-PRA-2009,GaKl-PRA-2011}, 
and tunnel ionization~\cite{KrXi-LaserPhys-1992,BrCo-PRL-2005,BlDi-NatPhys-2009,FlWo-PRL-2011,MuSp-PRL-2011,PfCi-NatPhys-2011}.

Recently, high-harmonic generation (HHG) has become a major tool in attosecond 
physics, allowing one to generate ultrashort light pulses with broad 
spectral bandwidths~\cite{CoKr-NatPhys-2007,LeLi-JPhysB-2008}.
From the ability to generate attosecond pulses~\cite{GoSc-Science-2008},
an entire new research area has emerged~\cite{KrIv-RMP-2009} focusing on 
electronic dynamics~
\cite{SmPa-PNAS-2009,SmIv-Nature-2009,Vr-Nature-2009,HaSa-NatPhys-2010,SmIv-NatPhys-2010,WoBe-Science-2011,HoSt-NatPhys-2011} 
and molecular motion~\cite{BaRo-Science-2006,Le-PRL-2005,WoBe-Nature466-2010,ZhRa-NatPhys-2012} 
on their fundamental time scale.
A particularly interesting aspect is the electron motion and the
corresponding hole creation dynamics during the ionization process~\cite{EcPf-Science-2008,PfCi-NatPhys-2012,WiGo-Science-2011}.
The high pulse intensities used in these experiments distort
significantly the potential of the electrons such that it is possible for the
electron to tunnel through or even travel over the barrier out of the system (i.e., tunnel 
ionization or barrier suppression regime, respectively).

A well-known model to describe tunnel ionization in atomic systems is the 
Ammosov, Delone, and Krainov (ADK) model~\cite{ADK-ZETF-1986,BiMa-AJPhys-2004}.
It applies to intense low-frequency fields, where a quasistatic approximation
can be made, meaning electrons follow adiabatically the external field. 
In these kind of fields, tunneling rates and final ion populations can be well reproduced by 
the ADK model~\cite{Sa-PRA-2002}.
Short, few-cycle pulses give access to instantaneous rather than cycle-averaged quantities and reveal the non-adiabatic behavior of the electronic motion.
In this case, an explicit time-dependent treatment of the ionization process is advantageous.
In combination with a multichannel theory, the dynamics in the relative phases between generated ionic states can be captured; something that cannot be done by the ADK model.
In the past, the state of the ion (after ionization) has not been of high interest, since it was not experimentally accessible.

Such a technique does now exist: attosecond transient absorption spectroscopy~\cite{GoKr-Nature-2010,SaYa-PRA-2011,GaBu-PRA-2011,BaLi-PRA-2012}.
Transient absorption spectroscopy has been used for years to study chemical reactions on the femtosecond time scale~\cite{DaRo-JCP-1987}.
However, just recently this technique has been extended to the attosecond 
regime~\cite{GoKr-Nature-2010}, where it is possible to probe the diagonal and off-diagonal elements of the ion density 
matrix (IDM) of the generated ion. 
From the off-diagonal IDM elements, the relative phase and the degree of 
coherence between the ionic states can be extracted, which are highly sensitive 
to the multichannel interactions occurring during the ionization process~
\cite{PaSa-PRL-2011}.
Attosecond transient absorption spectroscopy has also been used to study the dynamics of autoionizing 
states~\cite{WaCh-PRL-2010,LoGr-ChemPhys-2007,TaGr-PRA-2012,ChLi-PRA-2012} 
and to study the motion of an electron wave packet during ionization~\cite{HoSc-PRL-2011}.

The rapid technical advances in synthesizing light pulses made it possible to
generate sub-cycle NIR pulses lasting no longer than a few femtoseconds, 
and to reduce the jitter (time delay fluctuation) between the NIR pump pulse 
and the XUV probe pulse to tens of attoseconds \cite{WiGo-Science-2011}.
This time delay stability allows one to reliably probe the NIR-driven tunnel ionization 
dynamics within an optical cycle (approx. 2~fs) as a function 
of the pump-probe delay.
For non-overlapping pump and probe pulses, the transient absorption spectroscopy can be used to determine the instantaneous IDM at the time of the probe pulse~\cite{SaYa-PRA-2011}.

The aim of this study is to investigate the ion population dynamics in krypton 
within the pump pulse. 
We show that the instantaneous ionic state population can be well captured by
the transient absorption spectrum even for overlapping pump and probe pulses.
Furthermore, we observe strong modifications of the absorption lines in the transient absorption spectroscopy when pump and probe pulses have the maximum overlap. 

We show that these deformations can be understood by relative phase shifts in the ionic dipole.
We identify that the highly non-perturbative dressing of the $N$-electron states (particularly with the neutral ground state) is responsible for the phase shift.
Also the dressing of the ionic ($N-1$-electron) states, which leads to energy shifts in the ionic states, contributes to the phase shift.
The latter one is, however, much weaker than the first dressing mechanism.
Note that these two dressing mechanisms are quite different in nature.
The first mechanism dresses $N$-electron states and the second one dresses $N-1$-electron states.

To capture these dressing mechanisms during ionization, a description of the entire $N$-electron system is required.
We describe the dynamics of the full $N$-body wave function with a time-dependent configuration interaction singles (TDCIS) approach~\cite{GrSa-PRA-2010}.
The description of the pump and the probe steps of Ref.~\cite{GoKr-Nature-2010,WiGo-Science-2011} requires at least two active electrons, since the pump pulse ionizes an outer-valence electron and the probe pulse resonantly excites an inner-shell electron into the generated hole.
Therefore, it is crucial to use a multi-channel model, which goes beyond the single-active-electron (SAE) approximation, to describe the pump-probe process.

The paper is structured as follows.
In Sec.~\ref{sec:theory}, we give an overview of our TDCIS method, and describe the theory of attosecond transient absorption for overlapping pump and probe pulses.
The results are discussed in detail in Sec.~\ref{sec:results}.
In Sec.~\ref{sec:conclusion} we draw our conclusions.
Atomic units~\cite{MoNe-RMP-2008} are employed throughout, unless otherwise 
noted.

\section{Theoretical Methods}
\label{sec:theory}

\subsection{Equations of Motion}
\label{sec:theo_eom}
The time-dependent Schr\"odinger equation of an $N$-electron system exposed to 
linearly polarized electric fields is given by
\begin{subequations}
\begin{eqnarray}
\label{eq:SGL}
  i\frac{\partial}{\partial t} \ket{\Psi(t)}
  &=&
  \hat H(t) \ket{\Psi(t)},
\\
\label{eq:Hamiltonian}
  \hat H(t)
  &=&
  \hat H_0
  +
  \hat H_1
  -
  E(t) \hat z,
\end{eqnarray}
\end{subequations}
where $\ket{\Psi(t)}$ is the full $N$-electron wave function and $\hat H(t)$ is
the exact $N$-body Hamiltonian, which can be partitioned into three main parts:
(1) $\hat H_0 = \hat F - i \eta \hat W$ is the sum of the time-independent 
Fock operator $\hat F$ and a complex absorbing potential (CAP), which reads
$W(r) = [r-r_\text{CAP}]^2\,\Theta(r-r_\text{CAP})$, 
where $r$ is the radius, and $\Theta(r)$ is the Heaviside step function;
(2) the electron-electron interactions that cannot be captured by the mean-field
potential in $\hat H_0$ are captured by $\hat H_1$  
($=\hat V_C-\hat V_\text{HF}-E_\text{HF}$, for a detailed description of these
quantities see Refs. \cite{RoSa-PRA74-2006,GrSa-PRA-2010});
(3) the term $E(t)\,\hat z$ is the laser-matter interaction in the electric
dipole approximation using the length form. 
The CAP within $\hat H_0$ prevents artificial reflections of the ionized 
photoelectron from the radial grid boundary and is located far away from the 
atom such that all processes close to the atom are unaffected by the CAP. 
This is controlled by the parameter $r_\text{CAP}$.

By allowing only one electron to get excited or ionized out of the ground state
configuration, we strongly reduce the complexity of solving Eq.~\eqref{eq:SGL}.
A suitable way to achieve this goal is by exploiting the configuration interaction
(CI) language and describing the $N$-body wave function in terms of the
Hartree-Fock ground state $\ket{\Phi_0}$ and singly-excited configurations
$\ket{\Phi^a_i}$. This approximation is known as CI-Singles (CIS). 
The corresponding TDCIS $N$-electron wave function reads
\begin{subequations}
\begin{eqnarray}
\label{eq:wfct_ansatz}
  \ket{\Psi(t)}
  &=&
  \alpha_0(t) \ket{\Phi_0}
  +
  \sum_{i,a} \alpha^a_{i}(t) \ket{\Phi^a_{i}},
\\
\label{eq:1p1hstate}
  \ket{\Phi^a_{i}}
  &=&
  \hat c^\dagger_{a} \hat c_{i}
  \ket{\Phi_0},
\end{eqnarray}
\end{subequations}
where $i,j$ and $a,b$ refer to occupied orbitals and unoccupied (virtual) 
orbitals, respectively, in the Hartree-Fock ground state $\ket{\Phi_0}$.
Indicies $p,q$ are used when no distinction is made between occupied and 
virtual orbitals.
The operators $\hat c^\dagger_{p}$ and $\hat c_{p}$ create
and  annihilate, respectively, an electron in the spin orbital $\ket{\varphi_p}$.
The equations of motion (EOM) for the expansion coefficients $\alpha_0(t)$ and 
$\alpha^a_{i}(t)$ read:
\begin{subequations}
\label{eq:eom}
\begin{eqnarray}
\label{eq:eom_alpha0}
  i\dot\alpha_0(t)
  &=&
  -
  E(t) \sum_{i,a} 
  \sbra{\Phi_0}\hat z\sket{\Phi^a_{i}}
  \alpha^a_{i}(t) 
\\
\label{eq:eom_alpha_ai}
  i\dot\alpha^a_{i}(t)
  &=&
  (\varepsilon_a-\varepsilon_i) \, \alpha^a_{i}(t)
  +
  \sum_{b,j}
  \sbra{\Phi^a_{i}}\hat H_1\sket{\Phi^b_{j}}
  \alpha^b_{j}(t) 
\\&&\nonumber
  - 
  E(t)
  \Big(\!
  \sbra{\Phi^a_{i}}\hat z\sket{\Phi_0}
  \alpha_0(t)
  \! + \!
  \sum_{b,j}
  \sbra{\Phi^a_{i}}\hat z\sket{\Phi^b_{j}}
  \alpha^b_{j}(t) 
  \Big),
\end{eqnarray}
\end{subequations}
where $\varepsilon_{p}$ are the orbital energies of the orbitals
$\ket{\varphi_{p}}$, which are eigenstates of the modified, time-independent 
Fock operator, i.e., 
$\hat H_0 \ket{\varphi_{p}} = \varepsilon_{p} \ket{\varphi_{p}}$.
The operator $\hat H_1$ is the residual electron-electron interaction, which goes beyond the mean-field potential.
The parentheses $|\cdot)$ and $(\cdot|$ stand for the vector and dual vector
with respect to the symmetric inner product required because of the non-hermiticity of $\hat H_0$.
The dipole interaction between singly-excited configurations reduces to transitions between states of the excited electron and transitions between ionic states: 
\begin{align}
  \label{eq:efield_phph-transitions}
  \sbra{\Phi^a_{i}}\hat z\sket{\Phi^b_{j}}
  &=
  \sbra{\varphi_a}\hat z\sket{\varphi_b} \delta_{i,j}
  -
  \sbra{\varphi_j}\hat z\sket{\varphi_i} \delta_{a,b}
  .
\end{align}
A detailed description of our implementation of the TDCIS method can be found
in Ref.~\cite{GrSa-PRA-2010,PaGr-PRA-2012}.

From the full $N$-body wave function one can construct the ion density matrix 
(IDM) $\hat\rho^\IDM(t)$ by tracing over the excited electron. The matrix
elements are given by
\begin{align}
\label{eq:IDM}
  \rho^\IDM_{i,j}(t)
  =&
  \sum_{a,b} 
  \Big(
    \alpha^a_i(t) \left[\alpha^b_j(t)\right]^*
    o_{b,a}
    + 
    2\eta\ e^{i(\varepsilon_i-\varepsilon_j)t}
\\\nonumber &
   \times \int_{-\infty}^t \!\!\! dt'\
    w_{b,a}\ \alpha^a_i(t')\,[\alpha^b_j(t')]^*\,
    e^{-i(\varepsilon_i-\varepsilon_j)t'}
  \Big)
  ,
\end{align}
where $w_{b,a}$ are the matrix elements of the CAP in the virtual orbital 
basis, and $o_{b,a}$ are the overlap matrix elements between virtual orbitals.
The second term in Eq.~\eqref{eq:IDM} corrects the loss of norm in the
IDM due to the absorption of the excited electron by the CAP. 
The CAP is placed far away from the atom such that an electron so far
out does not affect the ion, specifically the ionic states.
Therefore, the absorption of an electron by the CAP results only in an 
artificial loss of norm that is compensated by the second term in 
Eq.~\eqref{eq:IDM}.
To keep the notation compact, we use the notation 
$\rho^\IDM_{i}(t):=\rho^\IDM_{i,i}(t)$ for ionic state populations.

\subsection{Spin-Orbit Splitting}
\label{sec:theo_LS}
In order to include the effect of spin-orbit splitting in the occupied orbitals,
we follow the logic of Ref. \cite{RoSa-PRA79-2009}, where we account for 
spin-orbit splitting with degenerate-state perturbation theory within 
the ($n,l$) orbital manifold. 
The occupied orbital $i$ is, then, characterized by the quantum numbers 
$n_i,l_i,j_i,m^J_i$, where $n_i$ is the principal quantum number, $l_i$ is the 
orbital angular momentum, $j_i$ is the total angular momentum, and $m^J_i$ is 
the projection of the total angular momentum onto the polarization direction of 
the external laser field. 
The occupied spin-orbitals read
\begin{eqnarray}
\label{eq:occorb_LS}
  \ket{\varphi_i}
  &=&
  \begin{pmatrix}
    C^{j_i,m^J_i}_{l_i,m^J_i-\half;s_i,\half}  & \ket{n_i,l_i,m^J_i-\half}  \\
    C^{j_i,m^J_i}_{l_i,m^J_i+\half;s_i,-\half} & \ket{n_i,l_i,m^J_i+\half}
  \end{pmatrix},
\end{eqnarray}
where the Clebsch-Gordan coefficient is given by 
$C^{l_3,m_3}_{l_1,m_1;l_2,m_2}=\braket{l_1,m_1;l_2,m_2}{l_3,m_3}$, 
$s_i=\half$ is the spin of the electron, and $\ket{n,l,m}$ are the 1-particle
orbitals obtained from a non-relativistic Hartree-Fock calculation.
The orbital energies $\varepsilon_i$ are taken from experimental ionization 
potentials.
For the virtual orbitals, we can neglect spin-orbital splitting and use the 
quantum numbers $n_a,l_a,\sigma_a,m^L_a$ to classify the orbitals, where
$\sigma_a$ is the spin component in the laser polarization direction and
$m^L_a$ is the projection of the orbital angular momentum onto the 
laser polarization direction.
The virtual orbitals read
\begin{eqnarray}
\label{eq:virtorb}
  \ket{\varphi_a}
  &=&
  \ket{n_a,l_a,m^L_a}
  \begin{pmatrix}
    \delta_{\sigma_a,\half} \\
    \delta_{\sigma_a,-\half} 
  \end{pmatrix}.
\end{eqnarray}

The usage of linearly polarized light leads to the condition, 
$m^J_i=m^L_a+\sigma_a$, for each singly-excited configuration $\ket{\Phi^a_i}$,
which is conserved by $\hat H(t)$.
After the introduction of spin-orbit splitting for the occupied orbitals, we 
cannot make use of the $\sigma$ and $m^L$ symmetries independently to reduce 
the number of singly-excited configurations, $\ket{\Phi^a_i}$, as done in 
Ref. \cite{PaGr-PRA-2012}.
However, not all symmetries are lost and we find that Eq.~\eqref{eq:eom_alpha_ai}
is (up to a global phase) invariant under the parity transformation 
$(m^J_i,m^L_a,\sigma_a)\rightarrow(-m^J_i,-m^L_a,-\sigma_a)$.
The new parity-adapted, singly-excited configurations $\ket{\Phi^a_{i}}_\pi$ 
read
\begin{eqnarray}
\label{eq:1p1h_symmetry}
  \ket{\Phi^a_{i}}_\pi
  &=&
  \sqrthalf
  \left[
  \Big.
    \ket{\Phi^a_{i}}_{+}
    +
    (-1)^{ l_i+s_i-j_i + \pi}
    \ket{\Phi^a_{i}}_{-}
  \right]
  ,
\end{eqnarray}
where the configurations $\ket{\Phi^a_{i}}_{\pm}$ stand for singly-excited 
configurations with $ m^J_i \gtrless 0$, respectively.
States with $\pi=0$ are {\it gerade} parity state and $\pi=1$ states are 
{\it ungerade} parity state.
All {\it ungerade} configurations $\ket{\Phi^a_{i}}_{\pi=1}$ will never get
populated and we can exclude them in our further investigations.
Note that the factor $(-1)^{l_i+s_i-j_i}$ comes from the symmetry
$C^{l_3,m_3}_{l_1,m_1;l_2,m_2} = (-1)^{l_1+l_2-l_3}\, 
C^{l_3,-m_3}_{l_1,-m_1;l_2,-m_2}$.

\subsection{Matrix Elements}
\label{sec:theo_matrix}
The matrix elements, which are needed for Eqs.~\eqref{eq:eom}, must be 
evaluated in the parity-adapted $\ket{\Phi^a_i}_{\pi}$ configuration basis.
The dipole matrix elements with respect to the symmetric inner product read
\begin{widetext}
\begin{subequations}
\label{eq:mat_z}
\begin{eqnarray}
\label{eq:mat_z_1p1h-1p1h}
  \big._{\pi_1} \!\!\sbra{\Phi^a_{i}} \hat z \sket{\Phi^b_{j}}_{\pi_2}
  &=&
  \left(
    \delta_{i,j}\,\delta_{\sigma_a,\sigma_b}\, z_{(A,B)}^{m^L_a} 
    -
    \delta_{a,b}\,\delta_{m^J_i,m^J_j}\, \sum_\sigma z_{(J,I)}^{m^J_i-\sigma} 
     C^{j_i,m^J_i}_{l_i,m^J_i-\sigma;s_i,\sigma} \
     C^{j_j,m^J_j}_{l_j,m^J_j-\sigma;s_j,\sigma} 
  \right)
\\\nonumber&&
  \times\,
  \delta_{\pi_1,\pi_2}
  ,
\\
\label{eq:mat_z_HF-1p1h}
  \sbra{\Phi_0} \hat z \sket{\Phi^a_{i}}_{\pi}
  &=&
  \delta_{\pi,0}\,\delta_{m^J_i,m^L_a+\sigma_a}\,
  \sqrt{2}
   \ 
  z_{(A,I)}^{m^L_a}\, C^{j_i,m^J_i}_{l_i,m^J_i-\sigma_a;s_i,\sigma_a},
\end{eqnarray}
\end{subequations}
where we used the notation $i=(I,j_i,m^J_i)$ with $I=(n_i,l_i)$ and 
$a=(A,m^L_a,\sigma_a)$ with $A=(n_a,l_a)$.
The dipole matrix elements in the original, non-parity-adapted basis are given 
by $z^{m}_{(P,Q)}:=\sbra{n_p,l_p,m}\hat z\sket{n_p,l_q,m}$.
The matrix elements of $\hat H_1$ read
\begin{eqnarray}
\label{eq:mat_coulomb}
  \big._{\pi_1} \!\!\sbra{\Phi^a_{i}} \hat H_1 \sket{\Phi^b_{j}}_{\pi_2}
  & = &
  2\,\delta_{\pi_1,0}\, \delta_{\pi_2,0}  \,
  C^{j_i,m^J_i}_{l_i,m^J_i-\sigma_a;s_i,\sigma_a} \
  C^{j_j,m^J_j}_{l_j,m^J_j-\sigma_b;s_j,\sigma_b} \ 
  v_{(AJIB)}^{M_1}
\\\nonumber&&  -
  \delta_{\pi_1,\pi_2}  \,
  \delta_{\sigma_a,\sigma_b}
  \sum_\sigma
  C^{j_i,m^J_i}_{l_i,m^J_i-\sigma;s_i,\sigma} \
  C^{j_j,m^J_j}_{l_j,m^J_j-\sigma;s_j,\sigma} \  
  v_{(AJBI)}^{M_1^\sigma}
\\\nonumber&&  -
  (-1)^{\pi_1}
  \delta_{\pi_1,\pi_2} \,
  \delta_{\sigma_a,-\sigma_b}
  \sum_\sigma
  C^{j_i,m^J_i}_{l_i,m^J_i-\sigma;s_i,\sigma} \
  C^{j_j,m^J_j}_{l_j,m^J_j+\sigma;s_j,-\sigma} \ 
  v_{(AJBI)}^{M_2^\sigma},
\end{eqnarray}
where ${M_1=(m^L_a,m^L_b,m^L_a,m^L_b)},\ 
{M_1^\sigma=(m^L_a,m^L_j-\sigma, m^L_b,m^J_i-\sigma)}$,
and ${M_2^\sigma=(m^L_a,-m^L_j-\sigma,-m^L_b,m^J_i-\sigma)}$.
The Coulomb matrix elements in the non-parity-adapted basis read
${v_{(PQRS)}^{M}\!\!:=\!\sbra{n_p,\!l_p,\!m_p;n_q,\!l_q,\!m_q} 
\!\hat r^{-1}_{12}\! \sket{n_r,\!l_r,\!m_r;n_s,\!l_s,\!m_s}}$ 
with ${M=(m_p,m_q,m_r,m_s)}$.

\end{widetext}

\subsection{Transient Absorption for Overlapping Pulses}
\label{sec:theo_tas}
The transient absorption signal is a direct measure of the cross section of the system.
In Ref. \cite{SaYa-PRA-2011} the transient absorption signal was derived for
the case of non-overlapping pump and probe pulses. 
The probe pulse was treated in first-order perturbation theory such that it 
was possible to give an analytic expression for the transient absorption 
signal as a function of the instantaneous IDM $\hat\rho^\IDM(t)$.
The pump pulse, usually a strong-field NIR pulse, which ionizes the atom by 
tunnel ionization, was treated non-perturbatively.

For overlapping pump and probe pulses, the influence of the probe pulse does not decouple from the impact of the pump pulse.
Therefore, it is not clear to which extent $\hat\rho^\IDM(t)$ can be extracted from the transition absorption spectrum like for non-overlapping pulses.
In order to fully capture the overall effect of pump and probe pulses, both pulses are treated non-perturbatively meaning the TDCIS equations of motion [see Eq.~\eqref{eq:eom}] are solved for an electric field $E(t)=E_\text{pump}(t)+E_\text{probe}(t)$.
Note that the probe step could also be treated perturbatively by introducing a two-time IDM, which depends on two different time arguments.
In our non-perturbative approach only the one-time IDM $\hat\rho^\IDM(t)$ needs to be constructed for each pump-probe configuration.
From $\hat\rho^\IDM(t)$ the ionic dipole moment 
\begin{eqnarray}
  \label{eq:dipole_ion}
  \left<z\right>_\text{ion}(t)
  &=& 
  \text{Tr}\left[\,\hat z\, \hat\rho^\IDM(t)\right]
\end{eqnarray}
and the atomic cross section 
\begin{align}
\label{eq:crosssection_atom}
  \sigma_a(\omega)
  & =
  4\pi\,\frac{\omega}{c}\,
  \text{Im}\left[
    \frac{ \left<z\right>_\text{ion}(\omega) }
    {E_\text{probe}(\omega)}
  \right]
\end{align}
can be calculated.
By performing the trace over $\hat\rho^\IDM(t)$ and not over the full $N$-body density matrix $\hat\rho(t)$, we consider only dipole transitions between ionic states.
Transitions between virtual orbitals can be neglected, since the XUV probe pulse interacts only weakly with the excited electron. 
Transitions between occupied and virtual orbitals describe stimulated emission and photoionization processes.
Both mechanism do not lead to sharp features in $\sigma_a(\omega)$ around the bound-bound transition energies.
Therefore, we ignore these contributions, which lead to background signals we are not interested in.

The detector, where the transient absorption spectrum is measured, does not record the atomic response but rather a damped spectrum of the form
\begin{align}
\label{eq:beers_law}
  \left| \big.
    E_\text{probe}(L,\omega)
  \right|^2
  & =
  \left| \big.
     E_\text{probe}(0,\omega)
  \right|^2 \
  e^{ -L\,n_\text{AT}\, \sigma_a(\omega)}
  ,
\end{align}
where $E_\text{probe}(0,\omega)$ is the incoming probe electric field, $L$ is the length of the medium, $E_\text{probe}(L,\omega)$ is the probe electric field at the end of the medium, and $n_\text{AT}$ is the atomic number density.
In Eq.~\eqref{eq:beers_law} Beer's law is used, which assumes a homogeneous medium and that the ratio $\left<z\right>_\text{ion}(\omega) / E_\text{probe}(\omega)$ is independent of $E_\text{probe}(\omega):=E_\text{probe}(0,\omega)$.
In Sec.~\ref{sec:res_tas} the validity of Beer's law is discussed.

Due to the finite energy resolution of the detector, the transient absorption signal in Eq.~\eqref{eq:beers_law} has to be convolved with a Gaussian mask function, where the full-width-at-half-maximum (FWHM) width is given by the energy resolution of the detector.
The cross section $\sigma_m(\omega)$ measured at the detector can be related to the atomic cross section and is given by
\begin{align}
\label{eq:crosssection_measured}
  \sigma_m(\omega)
  & =
  -\frac{1}{n_\text{AT}\,L}
  \ln\left(\Big.
    e^{
      -n_\text{AT}\,L\,\sigma_a(\omega)
    }
    *
    G_{\delta E}(\omega)
  \right)
  ,
\end{align}
where $G_{\delta E}(\omega)$ is the area-normalized Gaussian with the FWHM width
of $\delta E$, and the symbol $*$ stands for the frequency convolution.
The dependence on the pump-probe configuration, specifically the pump-probe delay $\tau$, enters parametrically in Eqs.~\eqref{eq:dipole_ion}-\eqref{eq:crosssection_measured} such that the atomic and the measured cross sections read $\sigma_a(\omega;\tau)$ and $\sigma_m(\omega;\tau)$, respectively.

\subsection{Oscillating Dipole Model}
\label{sec:theo_dipole}
In the following, we develop general expression for the transient absorption spectrum, which is based on a simple model.
Later in Sec.~\ref{sec:results} we use this generalized expression to discuss the features of the transient absorption spectrum for overlapping pulses obtained from our TDCIS calculations described in Sec.~\ref{sec:theo_eom}-\ref{sec:theo_matrix}.

First, we reduce the description of the ion to a two-level system.
The ground state $\ket{g}$ can only be accessed by the pump pulse via tunnel ionization and the excited state $\ket{e}$ can only be accessed by the probe pulse via resonant excitation out of $\ket{g}$.
The probe pulse, which may be approximated by a delta pulse, i.e., $E_\text{probe}(t;\tau)=E_0\,\delta(t-\tau)$, creates a coherent superposition $\ket{\Psi(t>\tau)}=a_0\ket{g}+a_1\, e^{-i(\omega_0 -i\Gamma/2)\,(t-\tau)} \ket{e}$, where $\omega_0$ is the positive energy difference between the two states, $1/\Gamma$ is the lifetime of the excited state, and $a_1=-iE_0\,a_0\, \bra{e}z\ket{g}$ results from the excitation by the probe pulse.
This superposition leads to an oscillating dipole 
\begin{align}
  \nonumber
  \label{eq:oscillating_dipole}
  \left<z\right>_\text{ion}(t>\tau)
  =&
  \bra{\Psi(t)} z \ket{\Psi(t)}
  =
  -2 E_0\,\big|\!\bra{e}z\ket{g}\!\big|^2 |a_0|^2\, 
\\
  &\times
  \sin[\omega_0\, (t-\tau)] \, 
  e^{-\frac{\Gamma}{2}\, (t-\tau)}
  .
\end{align}
Inserting Eq.~\eqref{eq:oscillating_dipole} in Eq.~\eqref{eq:crosssection_atom} and using $E_\text{probe}(\omega;\tau)=E_0e^{-i\omega\tau}$ the final expression for the cross section reads
\begin{eqnarray}
  \label{eq:cs_dipole_ion}
  \sigma(\omega;\tau)
  &=&
  \frac{4\pi\,\omega}{c}\,
  z_0
  \frac{
     \Gamma/2 
  }{
    (\omega-\omega_0)^2+\frac{\Gamma^2}{4}
  }
  ,
\end{eqnarray}
where $z_0=|a_0|^2 \, \big|\!\bra{e}z\ket{g}\!\big|^2$ determines the transition strength.
We see that for a simple two-level system the cross section is purely Lorentzian and directly proportional to the ground state population $|a_0|^2$ at the time of the probe step.

Adiabatic energy shifts in the ionic states during the intense NIR pulse result in a phase shift in the oscillating ionic dipole, i.e., $\expect{z}_\text{ion} \propto \sin[\omega_0(t-\tau)+\phi(\tau)]$.
Here, we assume the ionic state and the dipole oscillation live for a long time after the NIR pulse is over such that the entire dipole dynamics can be approximated by a phase-shifted oscillation.
The phase shift $\phi(\tau)$ has a dramatic influence on the shape of the transition line, which reads
\begin{align}
  \label{eq:cs_dipole_single}
  \sigma(\omega;\tau)
  &=
  \frac{4\pi\,\omega }{c}\,
  z_0 \
  \frac{
    \frac{\Gamma}{2}\,\cos[\phi(\tau)]
    + 
    (\omega-\omega_0)\,\sin[\phi(\tau)]
  }{
    (\omega-\omega_0)^2+\frac{\Gamma^2}{4}
  }
  .
\end{align}
Note that the phase shift $\phi(\tau)$ affects only the shape of the transition but not the strength $z_0$.

\begin{figure}[ht!]
\begin{center}
  \rmpdfinfo
  \includegraphics[width=\linewidth]{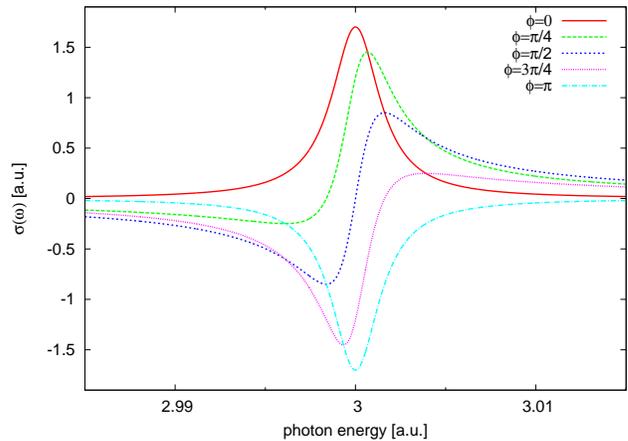}
  \caption{(color online) 
    The cross section $\sigma(\omega)$ of Eq.~\eqref{eq:cs_dipole_single} is shown for several values of $\phi(\tau)$.
    The transition energy is $\omega_0=3$~a.u., $\Gamma=3.2\times 10^{-3}$~a.u., and the transition strength is given by $z_0=0.01$.
  }
  \label{fig:cs_dipole-theo}
\end{center}
\end{figure}

In Fig.~\ref{fig:cs_dipole-theo} $\sigma(\omega;\tau)$ is shown for specific values of $\phi$. 
The transition line is purely Lorentzian for $\phi=0$.
In the case $\phi=\pi/2$, the cross section shows a dispersive behavior and has equally negative and positive regions that lie symmetrically around the field-free transition energy $\omega_0(0)$.
For all other phases, the cross section is a sum of these two scenarios and becomes asymmetric around $\omega_0$.
A phase shift by $\pi$ changes the sign of the cross section.
For $-\pi/2 \leq \phi \leq \pi/2$, the system shows an absorbing behavior whereas for $\pi/2 \leq \phi \leq 3/2\,\pi$ the system is rather emitting.
A similar scenario has been discussed in atomic helium, where neutral excited states are dressed by an IR pulse leading to emitting and absorbing patterns depending on the pump-probe delay~\cite{HeWe-arxiv-2012}.

Similarly to the dressing of the ionic states, the influence of the excited electron on the ion via the residual Coulomb interaction and via the pump field can lead to additional phase shifts in the oscillating dipole (see Sec.~\ref{sec:theo_dressing}).
Furthermore, corrections to the transition strength $z_0$ can occur, which may cause $z_0$ to be no longer directly proportional to the instantaneous hole population.
In order to capture these effects, we parametrize in addition to the phase $\phi(\tau)$ also the transition strengths $z_0\rightarrow z_0(\tau)$.
Our generalized version of Eq.~\eqref{eq:cs_dipole_single} for a multi-level ion reads
\begin{eqnarray}
  \label{eq:cs_dipole}
  \sigma_\text{dipole}(\omega;\tau)
  &=&
  \frac{4\pi\,\omega }{c}
  \sum_T
  z_T(\tau)\,
\\\nonumber
  &&\times
  \frac{
    \frac{\Gamma_T}{2}\,\cos[\phi_T(\tau)] 
    + 
    (\omega-\omega_T)\,\sin[\phi_T(\tau)]
  }{
    (\omega-\omega_T)^2+\frac{\Gamma_T^2}{4}
  }
  ,
\end{eqnarray}
where we sum over all possible ionic transitions $T$.
Note that Eq.~\eqref{eq:cs_dipole} is designed to capture the influence of all these effects that go beyond our simple two-level model [see Eq.~\eqref{eq:cs_dipole_ion}].
However, Eq.~\eqref{eq:cs_dipole} cannot explain why these changes occur and where they come from.

\subsection{Mechanisms leading to the Phase Shift}
\label{sec:theo_dressing}
As discussed in Sec.~\ref{sec:theo_dipole}, the dressing of the ion can induce a phase shift in the ionic dipole.
In the following, we discuss in the language of TDCIS how the dressing by the field and the coupling of the excited electron to the ionic subsystem can influence the phases $\phi_T(\tau)$ of Eq.~\eqref{eq:cs_dipole}. 
First, we analyze the scenario where the time evolutions of the excited electron ($a$ index) and the ionic states ($i$ index) are decoupled. 
This is the case when the terms $\sbra{\Phi^a_i}\hat H_1\sket{\Phi^b_j}$ and $\sbra{\Phi^a_i}\hat z\sket{\Phi_0}$ are switched off in Eq.~\eqref{eq:eom_alpha_ai}.
The resulting EOM can be written as
\begin{align}
  \label{eq:eom_simplified}
  i\dot\alpha^a_{i}(t)
  &=
  \sum_b 
    H^\text{elec}_{(a,b)}(t) \,  \alpha^b_{i}(t)
  +
  \sum_j
    H^\text{ion}_{(i,j)}(t) \,  \alpha^a_{j}(t)
  ,
\end{align}
where $H^\text{elec}_{(a,b)}(t):=\varepsilon_b \, \delta_{a,b} + E(t) \sbra{\varphi_a}\hat z\sket{\varphi_b}$ affects only the excited electron, and $H^\text{ion}_{(i,j)}(t):=-\varepsilon_j \, \delta_{i,j} - E(t) \sbra{\varphi_j}\hat z\sket{\varphi_i}$ affects only the ionic states.
$\hat H^\text{elec}$ and $\hat H^\text{ion}$ can be viewed as Hamiltonians of the two subsystems. 
Note that the Hamiltonians of both subsystems commute, i.e., $\left[ \hat H^\text{elec}(t),\hat H^\text{ion}(t)\right]=0$.
To confirm that Eq.~\eqref{eq:eom_simplified} leads to decoupled EOMs for the excited electron and the ion, we make the product ansatz $\alpha^a_i(t)=\chi_a(t)\kappa_i(t)$, where the EOMs of the separated electronic and ionic wave functions are given by 
\begin{subequations}
\label{eq:eom_decoupled}
\begin{align}
  \label{eq:eom_decoupled_a}
  i\dot\chi_a(t)
  &=
  \sum_b 
    H^\text{elec}_{(a,b)}(t) \,  \chi_b(t)
  ,
\\
  \label{eq:eom_decoupled_i}
  i\dot\kappa_i(t)
  &=
  \sum_j
    H^\text{ion}_{(i,j)}(t) \,  \kappa_{j}(t)
  .
\end{align}
\end{subequations}
We find that the product ansatz with the decoupled EOMs of Eq.~\eqref{eq:eom_decoupled} solves Eq.~\eqref{eq:eom_simplified}.
This shows that Eq.~\eqref{eq:eom_simplified} is the overall EOM of the full $N$-electron system, which consists of two totally separated subsystems.
Note that the term $\sbra{\Phi^a_i}\hat z\sket{\Phi^b_j}$ [cf. Eq.~\eqref{eq:efield_phph-transitions}] does affect the excited electron and the ion but it does not lead to interactions between the two.
Enforcing a normalized electron wave function, we find from Eq.~\eqref{eq:eom_decoupled_i} that the IDM is given by 
\begin{align}
  \label{eq:idm_decoupled_tau}
  \rho^\IDM_{i,j}(t) 
  &=  
  \kappa^*_j(t)\kappa_i(t)
  =
  e^{i\omega_T\,[t-\tau]}\,
  \rho^\IDM_{i,j}(\tau) \,
  e^{i \phi_T^\text{ion}(t,\tau)}
  ,
\end{align}
where $T=i\rightarrow j$ denotes the ionic transition.
If $E(t)=0$ for all times $t$, we find $\phi_T^\text{ion}(t,\tau)=0$ for all $t$ and $\tau$.
Hence, the additional phase $\phi_T^\text{ion}(t,\tau)$ enters only due to the term $-E(t) \sbra{\varphi_j}\hat z\sket{\varphi_i}$ in $\hat H^\text{ion}$.
This is exactly the field-driven dressing of the ionic system.
After the pulse is over [$E(t)=0$], the phase $\phi_T^\text{ion}(t,\tau)$ becomes independent of $t$ and depends only on the probe time $\tau$, i.e., $\phi_T^\text{ion}(t,\tau) \rightarrow \phi_T^\text{ion}(\tau)$.
In Sec.~\ref{sec:res_polarizability}, we show how the field-driven dressing of the ionic system can be analytically analyzed with the help of the polarizability of the ion.
It is interesting to note that the influence of the electric field on the excited electron, i.e., $E(t) \sbra{\varphi_a}\hat z\sket{\varphi_b}$, does not influence the IDM and subsequently the ionic dipole oscillation.

Additional phase shifts similar to $\phi_T^\text{ion}(\tau)$ can also occur due to the coupling between the ion and the excited electron.
There exist two kinds of mechanism that can couple these two subsystems: 
(1) the residual Coulomb interaction [the $\sbra{\Phi^a_i}\hat H_1\sket{\Phi^b_j}$ in Eq.~\eqref{eq:eom_alpha_ai} are the corresponding matrix elements], 
and (2) the field-driven mixing of the excited $N$-electron states with the neutral ground state [the $\sbra{\Phi^a_i}\hat z\sket{\Phi_0}$ in Eq.~\eqref{eq:eom_alpha_ai} are the corresponding matrix elements].
Both terms were ignored in Eq.~\eqref{eq:eom_simplified}, which led to two decoupled subsystems.
To distinguish the phase shifts induced by the two different mechanisms, we introduce $\phi_T^\text{residual}(\tau)$ and $\phi_T^\text{ground}(\tau)$.
 The phase shift due to the residual Coulomb interaction is denoted by $\phi_T^\text{residual}(\tau)$, and $\phi_T^\text{ground}(\tau)$ denotes the phase shift due to the field-driven mixing to the neutral ground state.

Adding up all three phase shifts we find that the total phase shift for the transition $T$ is given by
\begin{align}
  \label{eq:phase_contributions}
  \phi_{T}(\tau)
  &=
  \phi^\text{ion}_{T}(\tau)
  +
  \phi^\text{residual}_{T}(\tau)
  +
  \phi^\text{ground}_{T}(\tau)
  .
\end{align}
In Sec.~\ref{sec:res_phase} we discuss which of the three phase shifts gives the dominant contribution.


\section{Results}
\label{sec:results}

\subsection{System, Pulse, and Numerical Parameters}
\label{sec:res_para}
All presented results were calculated with the \textsc{xcid} package~\cite{xcid}.
All calculations were performed with 600 radial grid points,
a maximum radius $r_\text{max}=150$, and a non-linear grid mapping parameter 
$\zeta=1.0$.
The CAP starts at a radius $r_\text{CAP}=130$ and has a strength $\eta=0.003$.
The maximum angular momentum employed was $l_\text{max}=30$.
Furthermore, if the orbital angular momentum of any 1-particle orbital $\ket{\varphi_p}$ appearing in the matrix elements of $\hat H_1$ exceeded $4$, then only the monopole term of $\hat H_1$ was considered.
With these parameters, our results are converged.
A detailed explanation about the parameters of the grid, of the CAP, and of the residual 
Coulomb interaction can be found in Ref.~\cite{GrSa-PRA-2010}.

\begin{figure}[ht!]
\begin{center}
  \rmpdfinfo
  \includegraphics[width=\linewidth]{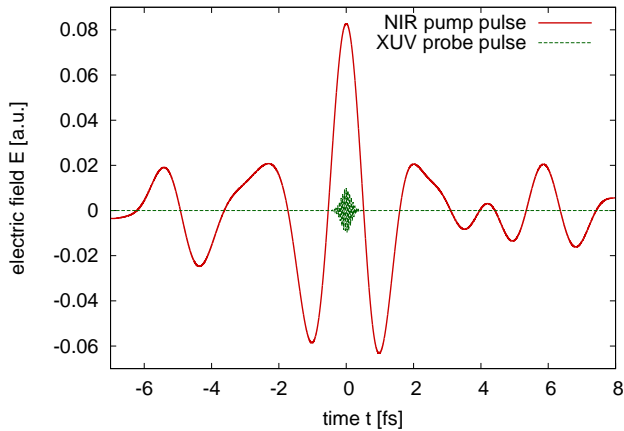}
  \caption{(color online) NIR pump (red solid) and XUV probe pulses (green dashed)
    used for transient absorption spectroscopy are shown separately. 
    The peaks of both pulses are centered at $t=0$.
  }
  \label{fig:efield}
\end{center}
\end{figure}

The NIR pulse profile measured in Ref.~\cite{WiGo-Science-2011} is used as the pump pulse in all the following results and is shown in Fig.~\ref{fig:efield}. 
The pump pulse is approx. 2~fs long with one main and two side peaks. 
The maximum electric field strength of the pump pulse is $\approx 0.08$~a.u.
(corresponding to an instantaneous intensity of $4.8\times10^{14}$~W/cm$^2$).
For the probe pulse (also shown in Fig.~\ref{fig:efield}) a Gaussian pulse 
profile is used with a central frequency of 3~a.u. ($\approx 81$~eV), a FWHM-width of the intensity profile of 10 a.u.($\approx 240$~as), a carrier envelope phase of zero, and a maximum field strength of $10^{-2}$~a.u. ($\approx 3.4$~TW/cm$^2$).

As in Refs.~\cite{GoKr-Nature-2010,WiGo-Science-2011}, we choose atomic krypton
as our system of interest.
The spin-orbit coupling within the occupied orbitals is accounted for in 
first-order perturbation theory as described in Sec. \ref{sec:theo_LS}.
The energies of the singly-ionized ionic states $[NL^M_J]^{-1}$ with respect 
to the Hartree-Fock ground state energy are given by the negative orbital 
energies (i.e., Koopmans' theorem).
Here, $N$ is the principal quantum number, $L$ the orbital angular momentum,
$J$ the total angular momentum, and $M$ the projection of $J$.
Since we are using linearly polarized light, the values for $M$ and $-M$ states are always the same.
Therefore, we refer always to a sum of both $|M|$ contributions, i.e., $[NL^M_J]^{-1}=[NL^{+M}_J]^{-1}+[NL^{-M}_J]^{-1}$.

The orbital energies of the $4p$ orbitals are replaced by experimental values
taken from Ref.~\cite{NISTwebsite} in order to match the experimental ionization potentials.
The orbital energies of the $4p$-shell are given by $\varepsilon_{4p_{1/2}}=-0.5389\ (=-14.67~\text{eV})$ and 
$\varepsilon_{4p_{3/2}}=-0.5148\ (=-14.00~\text{eV})$.
The orbital energies of the $3d$-shell are taken from Ref.~\cite{XDB}.
In addition, we account for the finite lifetime of all $3d^{-1}$ configurations of 7.5~fs ($\Gamma=88$~meV)~\cite{GoKr-Nature-2010}.
Hence, the $3d$ orbital energies become complex and read $\varepsilon_{3d_{3/2}}=-3.5-i\Gamma/2\ (=-95.24~\text{eV}-i0.044~\text{eV})$ and $\varepsilon_{3d_{5/2}}=-3.4525-i\Gamma/2\ (=-93.95~\text{eV} -i0.044~\text{eV})$.

\subsection{Transient Absorption Spectrum} 
\label{sec:res_tas}
In a recent experiment \cite{WiGo-Science-2011} attosecond transient absorption
spectroscopy has been used to investigate the hole production dynamics in 
atomic krypton during a sub-cycle NIR pump pulse. 
The theory described in Ref.~\cite{SaYa-PRA-2011} was used to analyze the
transient absorption spectrum and to connect the spectrum to the instantaneous 
IDM $\rho^\IDM_{i,j}(t)$. 
Propagation effects and the finite energy resolution of the detector were 
accounted for as described in Eq.~\eqref{eq:crosssection_measured}.
The macroscopic propagation is captured by Beer's law. 
Previous studies~\cite{SaYa-PRA-2011} have shown that Beer's law is valid for the pump-probe scenario investigated here. 
Similar conclusions have been found in Ref.~\cite{GaBu-PRA-2011} when the probe pulse is much shorter than the pump pulse (as in our case). 
For probe pulses longer than the pump pulse, the macroscopic propagation can quite strongly deviate from Beer's law.

The theory developed in Ref.~\cite{SaYa-PRA-2011} is strictly speaking
only valid when pump and probe pulses do not overlap and the tunnel-ionized 
electron is far away from the parent ion. 
In this case, the ionic subsystem, i.e., $\hat \rho^\IDM(t)$, reduces to a simple multi-level system without any kind of interactions with the environment and between levels.
Hence, the entire dynamics of the ion is analytically known and reads $\rho^\IDM_{i,j}(t)=\rho^\IDM_{i,j}(t_0) e^{i(\varepsilon_i-\varepsilon_j)(t-t_0)}$.
For overlapping pulses, the dynamics of $\hat \rho^\IDM(t)$ become more complex.
Therefore, it is not clear to which extent the transient absorption spectrum
for a given pump-probe delay can be related to $\rho^\IDM_{i,j}(t)$ when pump and
probe pulses do overlap.

\begin{figure*}[ht!]
\begin{center}
  \rmpdfinfo
  \includegraphics[width=\linewidth]{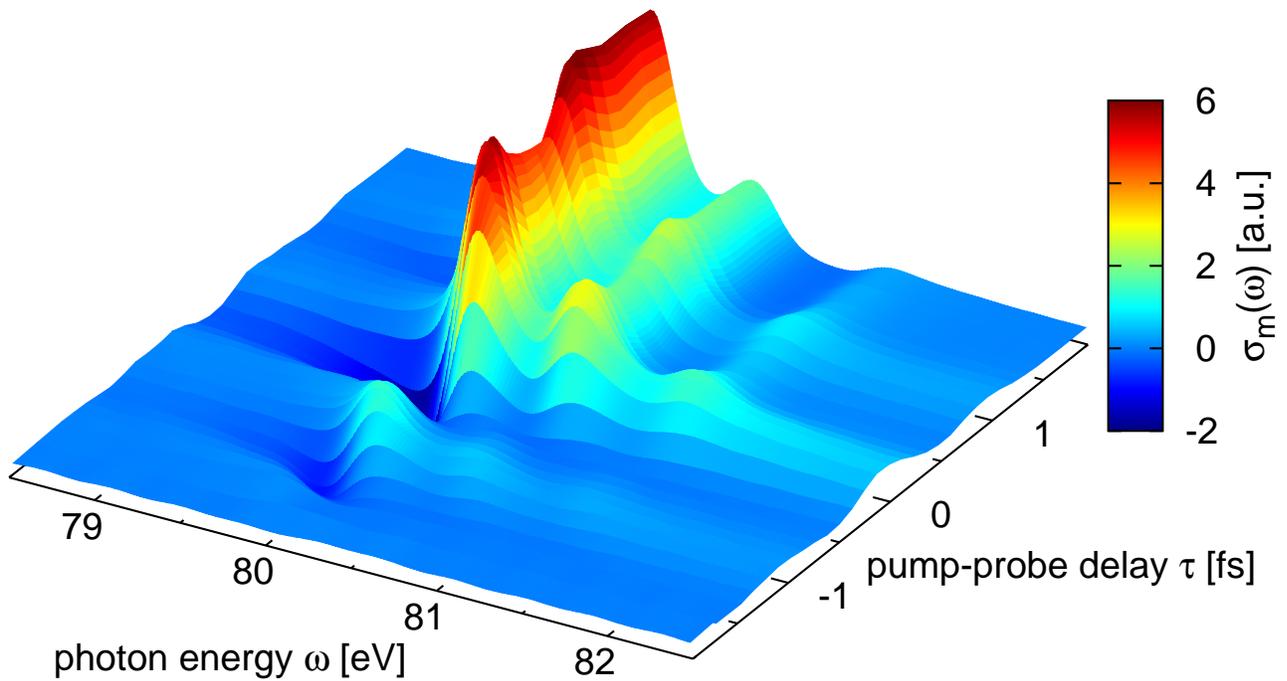}
  \caption{(color online) Attosecond XUV transient absorption spectrum 
    $\sigma_m(\omega;\tau)$ [see Eq. \eqref{eq:crosssection_measured}] of 
    krypton as a function of energy $\omega$ and pump probe delay $\tau$.
  }
  \label{fig:tas_3d}
\end{center}
\end{figure*}

In Fig.~\ref{fig:tas_3d} the calculated transient absorption spectrum $\sigma_m(\omega;\tau)$ 
is shown as a function of photon energy $\omega$ and pump-probe delay $\tau$.
The three main transition lines, i.e., $4p^{-1}_{3/2}\rightarrow 3d^{-1}_{5/2}$  
(79.95~eV), $4p^{-1}_{1/2}\rightarrow 3d^{-1}_{3/2}$ (80.57~eV),
and $4p^{-1}_{3/2}\rightarrow 3d^{-1}_{3/2}$ (81.24~eV), are clearly visible.
To shorten the notation we refer to these three transition lines as $T_1,T_2$, and $T_3$, respectively.

\begin{figure}[ht!]
\begin{center}
  \rmpdfinfo
  \includegraphics[width=\linewidth]{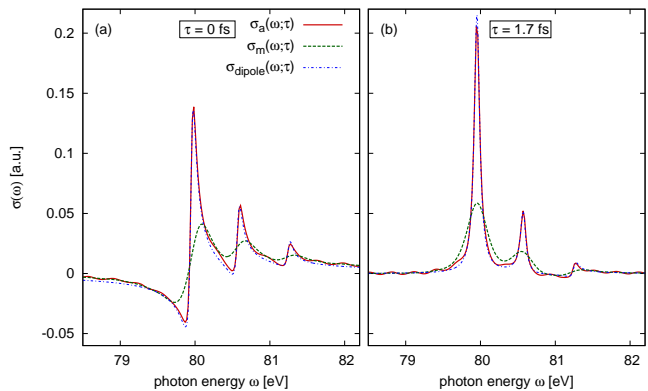}
  \caption{(color online) Transient absorption spectra for the pump-probe
    delay $\tau=0$ and $\tau=70\,(\approx 1.7~\text{fs})$ are shown in panel 
    (a) and (b), respectively.
    The atomic cross section $\sigma_a$ (red solid), the measured cross section 
    $\sigma_m$ (green dashed), and $\sigma_\text{dipole}$ (blue dotted) are 
    shown for each $\tau$.
  }
  \label{fig:cs_fixed-delay}
\end{center}
\end{figure}

The cross section shown in Fig.~\ref{fig:tas_3d} is in agreement with experimental observations~\cite{WiGo-Science-2011}.
The transition strengths increase mainly around $\tau \approx 0$, when the 
krypton atom is being probed within the main peak of the pump pulse.
It is also during this main peak of the pump pulse where the atom gets mainly ionized.
Simultaneously during the hole creation, the transition lines in the transient absorption spectrum change dramatically their shapes, resulting in negative cross sections for energies just below the field-free transition energies.
Negative cross sections can be only seen here for the $T_1$ transition, because a detector resolution of $\approx 300$~meV let these features disappear for the transitions $T_2$ and $T_3$ (cf. Fig.~\ref{fig:cs_fixed-delay}).

Similar to $\tau \approx 0$, line deformations and negative cross sections do also occur at $\omega\approx 80$~eV for $\tau \approx \pm1$~fs, where the probe pulse coincide with the side peaks of the pump pulse (cf. Fig.~\ref{fig:efield}).
The side peaks of the pump pulse are strong enough to lead to tunnel ionization as well (cf. Fig.~\ref{fig:pop}).
Particularly the ionization and the deformation caused by the first side peak of the pump pulse ($\tau \approx -1$~fs) can be nicely seen in Fig.~\ref{fig:tas_3d} for the transition line $T_1$.
As it will become clear in the discussion in Sec.~\ref{sec:res_phase}, the mechanism behind the deformations in all three cases ($\tau \approx 0,\pm 1$~fs) is the same.

For $\tau>2$~fs, the dynamics of the hole populations are barely affected by the pump pulse and behave as if they were field-free.
At these time delays, the intensity of the pump pulse is also strongly reduced (by more than a factor 10) compared to the peak intensity at $\tau=0$, thus supporting the observation of field-free behavior for larger pump-probe delays.
Field-free behavior means for the main transition line $T_1$ that it becomes stationary and does not change anymore in shape and strength.
The other two transition lines show interference effects from the coherent superposition of $4p^{-1}_{3/2}$ and $4p^{-1}_{1/2}$~\cite{SaYa-PRA-2011,GoKr-Nature-2010}.

Large negative time delays ($\tau<-2$~fs) are not of interest and, therefore, they are not shown in Fig.~\ref{fig:tas_3d} due to two reasons.
First, the ionic cross section [see Eq.~\eqref{eq:crosssection_atom} and Eq.~\eqref{eq:crosssection_measured}] is zero for large negative time delays, because only neutral krypton atoms exist prior to the NIR pulse.
Second, no electronic dynamics can be probed before the NIR pulse, since the neutral krypton atoms are in the electronic ground state.

In Fig.~\ref{fig:cs_fixed-delay} the transient absorption spectra 
$\sigma_a(\omega;\tau)$, $\sigma_m(\omega;\tau)$, and 
$\sigma_\text{dipole}(\omega;\tau)$ are shown for $\tau=0$ and 
$\tau=70~(\approx 1.7~\text{fs})$. 
The transition lines in $\sigma_m(\omega;\tau)$ are broadened with 
respect to the atomic cross section $\sigma_a(\omega;\tau)$ due to the 
propagation effect and the finite detector resolution ($\approx300$~meV), 
which is wider than the natural transition widths ($\Gamma_i=88$~meV).
The cross section $\sigma_\text{dipole}(\omega;\tau)$ is obtained by
fitting Eq.~\eqref{eq:cs_dipole} to $\sigma_a(\omega;\tau)$ obtained from the TDCIS calculations.
The energies $\omega_{T_i}$ and the natural widths $\Gamma_{T_i}$ of all transition lines $T_i$ are kept fixed (see Sec.~\ref{sec:res_para}) and only the magnitudes $z_{T_i}(\tau)$ and the phases $\phi_{T_i}(\tau)$ are fitted to $\sigma_a(\omega;\tau)$. 
The features of $\sigma_a(\omega;\tau)$ are well captured by $\sigma_\text{dipole}(\omega;\tau)$ for all $\tau$.
At $\tau=70$, the two strongest transition lines are Lorentzian-shaped as 
expected for non-overlapping pump and probe pulses.
The second ($T_2$) and especially the third ($T_3$) transition lines do not have a Lorentzian shape due to the coherent superposition of $4p^{-1}_{3/2}$ and $4p^{-1}_{1/2}$. 

The success of $\sigma_\text{dipole}(\omega;\tau)$ in capturing all features of $\sigma_a(\omega;\tau)$ shows that the influence of all terms in Eq.~\eqref{eq:eom}, which go beyond a simple two-level model (see Sec.~\ref{sec:theo_dressing}), can be understood by phase shifts $\phi_{T_i}(\tau)$ and changes in the oscillating dipole strengths $z_{T_i}(\tau)$.

However, the oscillating dipole model cannot explain what is the physical origin of $\phi_{T_i}(\tau)$ or whether or not $z_{T_i}(\tau)$ can be related to $\rho^\IDM(\tau)$.
The answers to these questions are discussed in the following.

\subsection{Population dynamics} 
\label{sec:res_pop}
\begin{figure}[ht!]
\begin{center}
  \rmpdfinfo
  \includegraphics[width=\linewidth]{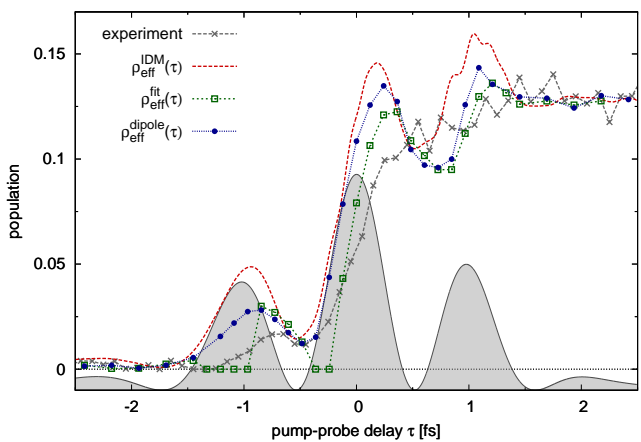}
  \caption{(color online) 
    The instantaneous hole population $\rho^\IDM_\text{eff}(\tau)$ (red solid) is shown together with the reconstructed populations $\rho^\text{dipole}_\text{eff}(\tau)$ (blue dotted) and $\rho^\text{fit}_\text{eff}$ (green dashed).
    The experimental population (grey dashed) is taken from Ref.~\cite{WiGo-Science-2011}.
    The reconstructed and experimental populations are scaled such that all have the same value at $t=2.4$~fs.
    The NIR pulse intensity is highlighted (gray area) in the background.
  }
  \label{fig:pop}
\end{center}
\end{figure}

First, we turn our focus to the population dynamics of the ionic states.
Particularly, we investigate the hole creation dynamics in the $4p_{3/2}$ orbitals during the pump pulse. 
In order to do so, we need to focus only on the main transition $T_1$.
For non-overlapping pulses, the transition strength is proportional to the instantaneous population
$\rho^\IDM_\text{eff}(\tau)
:=
\rho^\IDM_{4p^{1/2}_{3/2}}(\tau)
+
\frac{2}{3} \rho^\IDM_{4p^{3/2}_{3/2}}(\tau)
,
$
where $\rho^\IDM_{i}(\tau)$ are the hole populations of the ionic states $i=[4p^{1/2}_{3/2}]^{-1}$ and $i=[4p^{3/2}_{3/2}]^{-1}$ (for details on the notation see Sec.~\ref{sec:res_para}), respectively.
In Fig.~\ref{fig:pop}, we compare $\rho^\IDM_\text{eff}(\tau)$ (red solid) with the transition strength $\rho^\text{dipole}_\text{eff}:=z_{T_1}(\tau)$ (blue dotted), with the reconstructed population $\rho^\text{fit}_\text{eff}$ (green dashed) obtained by applying the same fitting procedure as in Ref.~\cite{WiGo-Science-2011} to $\sigma_m(\omega;\tau)$, and with the experimental data of Ref.~\cite{WiGo-Science-2011}.

In all three theoreticcal curves, oscillatory behavior can be seen during build up of the hole populations.
The features of $\rho^\IDM_\text{eff}(\tau)$ are well captured by $\rho^\text{dipole}_\text{eff}(\tau)$ and by $\rho^\text{fit}_\text{eff}(\tau)$.
It shows that $\rho^\IDM_\text{eff}(\tau)$ can be quite well reconstructed from the transient absorption spectrum even though pump and probe pulses do strongly overlap.
Note, however, that at the major peak ($\tau\approx 0$) a delay of about $\approx 200$~as occurs in $\rho^\text{fit}_\text{eff}(\tau)$.
The deviations from $\rho^\IDM_\text{eff}(\tau)$ are a measure of the influence of the excited electron on the ionic states, since the adiabatic dressing of the ionic states does not affect the transition strength.
As described in Sec.~\ref{sec:theo_dipole}, if the evolution of the excited electron decouples directly after the probe step from the evolution of the ion, $\rho^\text{dipole}_\text{eff}(\tau)$ coincides with $\rho^\IDM_\text{eff}(\tau)$.

In the experiment, this oscillatory behavior was not seen and the transition strengths increased monotonically with $\tau$.
The reason for this discrepancy might lie in the strong restriction of the CIS space for the ionic degree of freedom. 
Within the CIS space generated from the neutral ground state of the atom, the ion is described exclusively by one-hole configurations.

\subsection{Line Deformations and Phase Shifts} 
\label{sec:res_phase}
As we have seen in Sec.~\ref{sec:res_tas} the oscillating dipole model is able 
to describe all features of the transient absorption spectra by only adjusting 
the strengths $z_T(\tau)$ and the phases $\phi_T(\tau)$ of the three transitions.
The strengths $z_T(\tau)$ are in close connection to the instantaneous populations of 
the ionic states even for overlapping pulses [cf. Sec.~\ref{sec:res_pop}].
The correct values of $\phi_T(\tau)$ are important to capture the shapes of the transition lines, which can change significantly during the pump pulse [cf. Fig.~\ref{fig:cs_fixed-delay}].
\begin{figure}[ht!]
\begin{center}
  \rmpdfinfo
  \includegraphics[width=\linewidth]{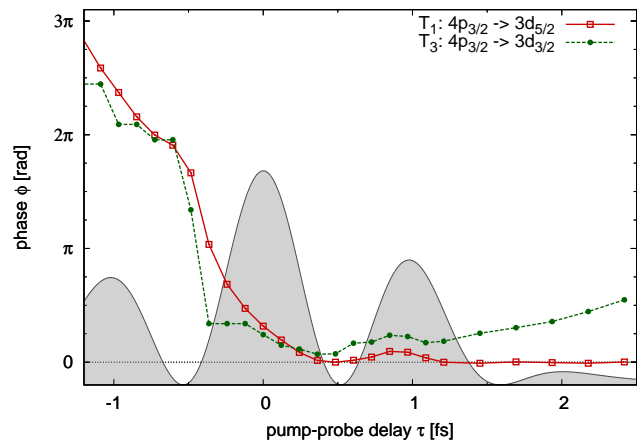}
  \caption{(color online) The phase shift $\phi_{T_1}(\tau)$ of the strongest transition line and the phase shift $\phi_{T_3}(\tau)$ of the weakest transition line are shown.
  }
  \label{fig:res_phase}
\end{center}
\end{figure}

In Fig.~\ref{fig:res_phase}, $\phi_{T_1}(\tau)$ and $\phi_{T_3}(\tau)$ are shown. 
The phase $\phi_T(\tau)$ is chosen such that the strengths $z_T$ are always positive.
The region $\tau<-1.5$~fs is not shown, since the strengths of the transitions 
are of the same order as the numerical background noise, which leads to large 
uncertainties in $\phi_T$.
The last peak of the pump field at $\tau=1$~fs has only a small effect on $\phi_T$ whereas the first peak at $\tau=-1$~fs has a strong influence on the phases.
This is an interesting observation, since both peaks have the same intensity.

As mentioned earlier, the coherent superposition of the ionic
states $4p^{-1}_{1/2}$ and $4p^{-1}_{3/2}$ results in a phase shift $(\varepsilon_{4p_{3/2}}-\varepsilon_{4p_{1/2}})\tau$, which is particularly dominant in transition $T_3$~\cite{SaYa-PRA-2011} and has been observed in previous experiments~\cite{GoKr-Nature-2010}.
This coherence-induced phase shift is also visible in $\phi_{T_3}(\tau)$ for $\tau>1$~fs (see Fig.~\ref{fig:res_phase}).
The slope of $\phi_{T_3}(\tau)$ is approx. $0.023$~a.u. which is in good agreement with $\varepsilon_{4p_{3/2}}-\varepsilon_{4p_{1/2}}=0.024$.
The main transition line $T_1$ is not affected by this coherent superposition and, therefore, goes over into a Lorentzian shape (i.e., $\phi_{T_1}=0$) for large $\tau$ (cf. Fig.~\ref{fig:res_phase}).

We have seen in Fig.~\ref{fig:res_phase} that the phase shift $\phi_T$ can be quite large.
However, what has not been answered yet is the origin of $\phi_T$.
As discussed in Sec.~\ref{sec:theo_dressing}, there are three main contributions to $\phi_T(\tau)$ [cf. Eq.~\eqref{eq:phase_contributions}]:
(1) $\phi_T^\text{ion}(\tau)$ induced by the NIR-driven dressing of the ionic system [the $\sbra{\varphi_j}\hat z\sket{\varphi_i}$ are the corresponding matrix elements];
(2) $\phi_T^\text{residual}(\tau)$ induced by the residual Coulomb interaction; 
and (3) $\phi_T^\text{ground}(\tau)$ induced by the NIR-driven mixing of the excited $N$-electron system with the neutral ground state [the $\sbra{\Phi^a_i}\hat z\sket{\Phi_0}$ are the corresponding matrix elements].
Particularly, in order to account for $\phi_T^\text{residual}(\tau)$ and $\phi_T^\text{ground}(\tau)$, it is important to have a multi-electron picture, which can describe the degrees of freedom of the ionized electron and of the ion.

\begin{figure}[ht!]
\begin{center}
  \rmpdfinfo
  \includegraphics[width=\linewidth]{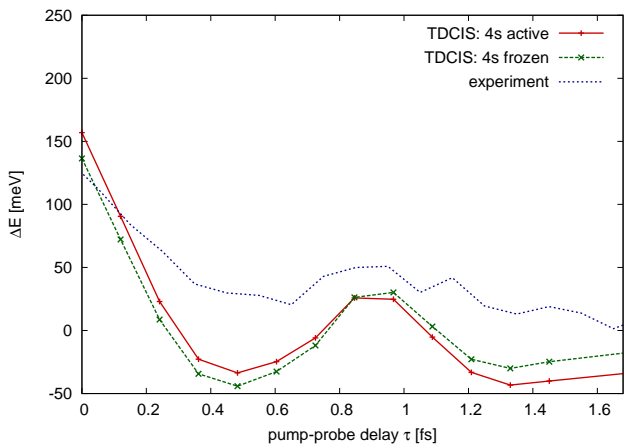}
  \caption{(color online) The apparent energy shifts of the strongest transition line $T_1$ are shown for the calculated cross sections with (red line with crosses) and without (green with asterisks) the $4s$ orbital active, and for the experimentally obtained cross sections (blue dotted).
    The experimental values are taken from Ref.~\cite{WiGo-Science-2011}.
  }
  \label{fig:energy-shift}
\end{center}
\end{figure}

The strong line deformations, if only the positive part of the cross section is considered, appear as if the transition energies have moved.
This energy shift we call apparent energy shift.
In Fig.~\ref{fig:energy-shift}, we make a direct comparison between calculated (red and green lines) and experimentally obtained (blue dashed line)~\cite{WiGo-Science-2011} energy shifts of the strongest transition line, $T_1$.
In the calculations yielding the red line (with crosses), no approximation is made in the TDCIS calculations.
For the green line (with asterisks), all couplings to the $4s$ orbital are turned off.
The procedure used to extract the apparent energy shift is described in Ref.\cite{WiGo-Science-2011}.

The magnitude of the apparent energy shifts are correctly reproduced by the calculations.
Ignoring the $4s$ orbitals makes no significant difference.
Particularly in the theoretical results, the second pump peak is clearly visible. 
For $\tau>1.4$~fs, the energy shifts go as expected to zero indicating that influence of the pulse decreases and the krypton ion can be treated as a field-free ion.
Negative pump-probe delays are not shown because the transition strengths are too weak to obtain reliable results.

In the following, we discuss each mechanism which can lead to the strong line deformations seen in Fig.~\ref{fig:tas_3d}.
We discuss each mechanism in terms of $\phi_T(\tau)$, since the oscillating dipole model showed that these deformations can be fully understood by a phase shift.

\subsubsection{Field Dressing of the Ionic System and the Polarizability of Kr$^+$}
\label{sec:res_polarizability}
In the presence of a strong external field, the energies of the ionic state become modified, i.e., $\varepsilon_i\rightarrow \varepsilon_i(E)$.
If the photon energy is much smaller than the ionization potential, the influence of the field on the ionic state can be described adiabatically.
This is the case for Kr$^+$, where the ionization potential is 24~eV~\cite{SuMu-JPCRD-1991}, and the photon energy of the pump pulse is 1.4~eV and, therefore, far off-resonance with any ionic transition.
Hence, the energy corrections $\Delta \varepsilon_i(E)=\varepsilon_i(E)-\varepsilon_i(0)$ of the dressed ionic states $i$ are well captured by the quadratic Stark effect
\begin{align}
  \label{eq:ionicstate_dressed}
  \Delta \varepsilon_i[E_\text{pump}(t)]
  =
  \frac{\alpha(i)}{2}\, E_\text{pump}^2(t)
  ,
\end{align}
where $\alpha(i)$ is the polarizability of the ionic state $i$, and the pump electric field is linearly polarized.
Note that the energy of the ionic state $i$ is $-\varepsilon_i$.
After the pulse is over, $\Delta \varepsilon_i(0)=0$ and the ion is back in its field-free state with the ionic energies $\varepsilon_i(0)$. 
If an oscillating dipole (coherent superposition of two states $i$ and $j$) is present while the energy of the ionic states get shifted, the dipole oscillates after the pulse with the field-free transition energy $\omega_{i \rightarrow j}(0)=\varepsilon_i(0)-\varepsilon_j(0)$ but phase-shifted by~\cite{WiGo-Science-2011}
\begin{subequations}
\label{eq:ionic_dressing}
\begin{align}
  \label{eq:stark_phase-shift}
  \phi^\text{ion}_{i \rightarrow j}(t)
  =&
  \int^{\infty}_t\!\!\! dt'\ 
    \Delta \omega_{i \rightarrow j}[E_\text{pump}(t')],
\\
  \label{eq:transition-energy-shift}
  \Delta \omega_{i  \rightarrow j}(E)
  =&
  \Delta \varepsilon_i(E) - \Delta \varepsilon_j(E)
  =
  \frac{\alpha(i)-\alpha(j)}{2}\
  E^2
  ,
\end{align}
\end{subequations}
where $\Delta \omega_{i  \rightarrow j}[E_\text{pump}(t)]$ is the instantaneous shift in the transition energy between the states $i$ and $j$.
The phase shift $\phi^\text{ion}_{i \rightarrow j}(t)$ is exactly the ionic phase shift $\phi_T^\text{ion}(\tau)$ with $T=i\rightarrow j$, which originates from the field-dressing of teh ionic system.

In order to estimate $\phi^\text{ion}_{T}(t)$ we use the quantum chemistry code {\sc dalton}~\cite{dalton} to calculate exact polarizabilities of the ionic states.
Furthermore, we can clarify to which extent our TDCIS calculation can correctly describe the polarizability of the ion and subsequently $\phi^\text{ion}_{i \rightarrow j}(t)$.
The polarizabilities were calculated with a complete active space self-consistent field (CASSCF) wave function and with a CIS wave function.
Here, CIS means that Kr$^+$ is described in the space of one-hole configurations, which is consistent with out TDCIS configuration space.
In CASSCF calculations, spatial deformations of the ionic states are included that would require CISD and higher order configuration excitations with respect to the ground configuration of the neutral atom.
These deformations are not included in CIS.
\begin{table}
  \caption{\label{tab:polar} Static dipole polarizabilities $\alpha_{x,x}=\alpha_{y,y}$ 
    and $\alpha_{z,z}$ of several states of Kr$^+$ are shown.
    Polarizabilities obtained by the CIS and CASSCF methods are compared.
    The CIS and CASSCF calculations are done with \textsc{dalton}.
    All values are given in atomic units and with a precision up to the 
    second digit.
  }  
  \begin{ruledtabular}
  \begin{tabular}{ l | c c | c  c }
    & \multicolumn{2}{c|}{CIS} & \multicolumn{2}{c}{CASSCF}  \\
    & $\alpha_{x,x}$ & $\alpha_{z,z}$  &  $\alpha_{x,x}$ & $\alpha_{z,z}$ \\
    \hline
    $[4p_{3/2}^{3/2}]^{-1}$   &  1.57  &  0.01 & 10.62 & 10.77   \\  
    $[4p_{3/2}^{1/2}]^{-1}$   &  0.53  &  2.09 & 10.72 & 10.57   \\ \hline
    $[4p_{1/2}^{1/2}]^{-1}$   &  1.05  &  1.05 & 10.67 & 10.67   \\ \hline
    $[3d_{5/2}^{5/2}]^{-1}$   & -0.01  &  0.00 &  9.57 &  9.71   \\ 
    $[3d_{5/2}^{3/2}]^{-1}$   &  0.00  &  0.00 &  9.63 &  9.60   \\ 
    $[3d_{5/2}^{1/2}]^{-1}$   &  0.00  & -0.01 &  9.65 &  9.54   \\ \hline
    $[3d_{3/2}^{3/2}]^{-1}$   &  0.00  &  0.00 &  9.58 &  9.68   \\ 
    $[3d_{3/2}^{1/2}]^{-1}$   &  0.00  & -0.01 &  9.65 &  9.55   \\ 
  \end{tabular}
  \end{ruledtabular}
\end{table}

In Table~\ref{tab:polar}, we summarize the results of the static dipole polarizabilities for several states of Kr$^+$ with holes in the $4p$ or in the $3d$ orbital manifolds. 
For more details about the polarizability calculations see Appendix~\ref{app:polar}.
The CASSCF calculations are in good agreement ($\pm2$\% accuracy) with the 
static polarizabilities in Ref.~\cite{medved}.
Polarizabilities obtained with the CASSCF method are in very good agreement ($<2\%$) with experimental results~\cite{KuMe-CanJChem-1985,HoKe-MolPhys-1990} for neutral krypton atoms.
For ionic krypton the correlation effects are reduced in comparison to neutral krypton, thus making the CASSCF calculations even more accurate.
Hence, we may assume that the CASSCF results for Kr$^+$ are practically exact.

The polarizabilities obtained with CASSCF have values around 10.7 for 
$4p^{-1}_j$ ionic states and values around 9.6 for $3d^{-1}_j$ ionic states.
In both cases, the anisotropy ($\alpha_{z,z}-\alpha_{x,x}$) is small.
This stands in contrast to the CIS results, where a high anisotropy in the
polarizabilities of $4p^{-1}$ is found.
Furthermore, the polarizabilities of the ionic states $3d^{-1}$ are basically zero with the CIS basis set.
The only contribution to $\alpha(3d^{-1})$ comes from the weak coupling between the $3d$ and $4p$ orbitals.
Within CIS, the polarizabilities for $4p^{-1}$ are determined by the coupling to the $4s^{-1}$ state.
Ionic states with two or more holes in the initially occupied orbitals do not exist in CIS and, therefore, cannot contribute to the polarizabilities.
This restriction in the CIS space limits the ability of the ionic states to respond to the external field and leads to much smaller polarizabilities (as seen in Table~\ref{tab:polar}).

The quantity that determines $\phi^\text{ion}_{T}(t)$ is the difference in the polarizabilities, not the polarizabilities themselves [see Eq.~\eqref{eq:ionic_dressing}].
We are only interested in the $\alpha_{zz}$ component, since we use light linearly polarized along the $z$ axis.
For the strongest transition line $4p^{-1}_{3/2}\rightarrow 3d^{-1}_{5/2}$, we need to look at the differences $\alpha([4p^m_{3/2}]^{-1})-\alpha([3d^m_{5/2}]^{-1})$ for the ionic states with $m=1/2$ and $m=3/2$.
\begin{table}
  \caption{\label{tab:polar_diff} Differences between the static dipole 
    polarizabilities $\alpha_{z,z}$ of the states involved in the 
    main transition line $4p^{-1}_{3/2}\rightarrow 3d^{-1}_{5/2}$.
    The values are given in atomic units.
  }  
  \begin{ruledtabular}
  \begin{tabular}{ c | c |  c }
    $\alpha([4p^m_{3/2}]^{-1})-\alpha([3d^m_{5/2}]^{-1})$ & CIS & CASSCF  \\ \hline
    $m=\half$       &  2.10  & 1.03   \\ 
    $m=\frac{3}{2}$ &  0.01  & 1.17
  \end{tabular}
  \end{ruledtabular}
\end{table}

In Table~\ref{tab:polar_diff} theses differences are shown.
The strong $m$-dependence of the CIS results is due to the high anisotropy of the polarizabilities.
The accurate CASSCF results show almost no dependence on $m$.
Since the $[4p_{3/2}^{m=3/2}]^{-1}$ population is in our calculation much smaller than the $[4p_{3/2}^{1/2}]^{-1}$ population, we focus only on the $m=\half$ results.
Contrary to the results of the polarizabilities, where the CIS approach underestimates the values, the difference $\alpha([4p^{1/2}_{3/2}]^{-1})-\alpha([3d^{1/2}_{5/2}]^{-1})$ is overestimated by CIS.
The maximum energy shifts in the transition for the given pump parameters
[cf. Fig.~\ref{fig:efield}] are 183~meV (CIS) and 102~meV (CASSCF), respectively.
Assuming the maximum energy shifts persist over 2~fs, the resulting phase shifts are 0.56~rad (CIS) and 0.31~rad (CASSCF), respectively.
This approximation can be nicely verified by a 3-level Bloch model describing only the ($N-1$)-electron ionic states $3d^{-1},4s^{-1}$, and $4p^{-1}$.

\begin{figure}[ht!]
\begin{center}
  \rmpdfinfo
  \includegraphics[width=\linewidth]{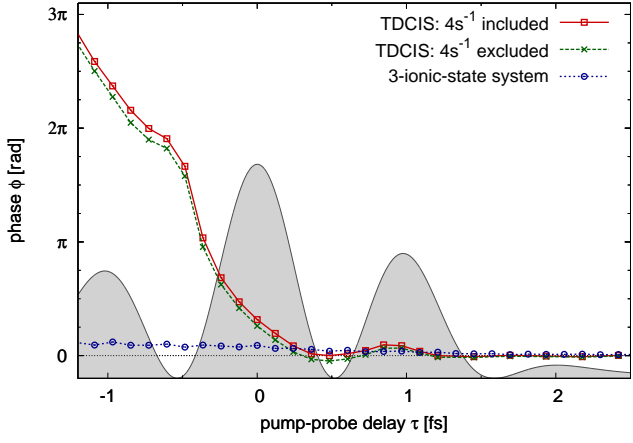}
  \caption{(color online) 
     The phase shifts $\phi_{T_1}(\tau)$ of the transition line $T_1 (4p^{-1}_{3/2} \rightarrow 3d^{-1}_{5/2})$ are shown for TDCIS calculations with (red solid) and without (greeen dashed) the ionic state $4s^{-1}$.
     Additionally, the phase shift $\phi^\text{ion}_{T_1}(\tau)$ (blue dotted) is shown for a Bloch model describing 3 ionic states (see text for details).
     Here, $\phi^\text{ion}_{T_1}(\tau)$ originates solely from the dressing of the ionic state $[4p^{1/2}_{3/2}]^{-1}$ through the coupling to $4s^{-1}$. 
  }
  \label{fig:phase_dressed-ion}
\end{center}
\end{figure}

The phase shift $\phi^\text{ion}_{T_1}$ is, however, much smaller than $\phi_{T_1}$ [see Fig.~\ref{fig:phase_dressed-ion}].
Hence, $\phi^\text{ion}_{T_1}(t)$ cannot be the main contribution to $\phi_{T_1}(t)$.
Furthermore, when the coupling to the $4s^{-1}$ ionic state is switched off and $\alpha(i)\approx0$ within TDCIS the phase shift $\phi_{T_1}(t)$ is almost unchanged. 
Note that in this case, the ionic states cannot be dressed and $\phi^\text{ion}_{T_1}=0$.

\subsubsection{Residual Coulomb interaction}
\label{sec:res_residual}
\begin{figure}[ht!]
\begin{center}
  \rmpdfinfo
  \includegraphics[width=\linewidth]{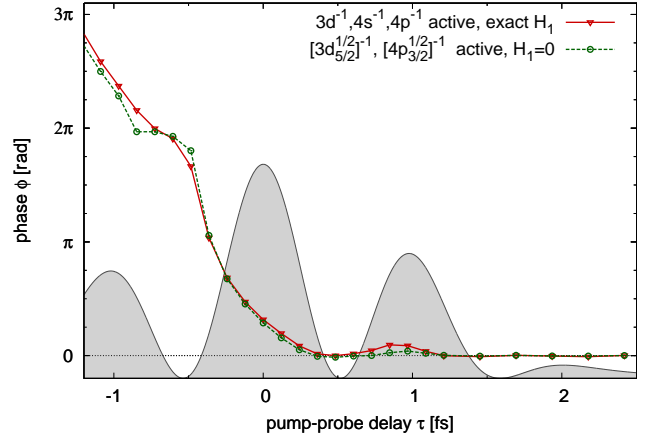}
  \caption{(color online)
      The phase shifts $\phi_{T_1}(\tau)$ are shown. 
      The TDCIS calculations were done on the one side (red solid) with the full electron-ion interaction and all $3d^{-1}, 4s^{-1}, 4p^{-1}$ ionic states active, and on the other side (green dashed) with a simplified two-level TDCIS model without residual Coulomb interaction ($\hat H_1=0$).
  }
  \label{fig:phase_residual}
\end{center}
\end{figure}

Here, we discuss the influence of the residual Coulomb interaction---specifically $\phi_T^\text{residual}(\tau)$.
First, we simplify the krypton atom to a two-level system with the states 
$[4p^{1/2}_{3/2}]^{-1}$ and $[3d^{1/2}_{5/2}]^{-1}$ such that no dressing of the ionic states can occur ($\phi_T^\text{ion}=0$).
The total phase shift reads now $\phi_{T}(\tau) = \phi^\text{residual}_{T}(\tau) + \phi^\text{ground}_{T}(\tau)$.
Second, we also ignore $\hat H_1$ such that the ionized electron can have no influence on the remaining ion via $\hat H_1$---this means we set $\phi_T^\text{residual}=0$.
In Fig.~\ref{fig:phase_residual}, we compare $\phi_{T_1}(\tau)$ obtained from the full TDCIS model including $\hat H_1$ and all ionic states of Kr$^+$ (red solid) and that obtained from the simplified two-channel TDCIS model (green dashed) just described.
Whether or not $\hat H_1$ is included makes no significant difference in the behavior of $\phi_{T_1}(\tau)$.  
Hence, residual Coulomb interaction between the ionized electron and the ion has almost no effect, i.e., $\phi_T^\text{residual}(\tau)\approx 0$.

\subsubsection{Field-Induced Mixing with the Neutral Ground State}
\label{sec:res_tunnel}
The discussions in Secs.~\ref{sec:res_polarizability} and \ref{sec:res_residual} have shown that $\phi_T^\text{ion} \ll \phi_T$ and $\phi_T^\text{residual}\ll \phi_T$.
Hence, we must conclude that the main reason for the phase shift comes from the field-driven mixing with neutral ground state, i.e., $\phi_T(\tau) \approx \phi_T^\text{ground}(\tau)$ [cf. Eq.~\eqref{eq:phase_contributions}].
Remember that the mixing to the ground state is captured by the terms $\sbra{\Phi_0}\hat z\sket{\Phi^a_i},\sbra{\Phi^a_i}\hat z\sket{\Phi_0}$ [cf. Eq.~\eqref{eq:eom}].
These terms are also responsible for describing tunnel ionization.
To verify that the field-induced mixing of the excited $N$-electron states $\Phi^a_i$ with $\Phi_0$ is indeed the main reason for the observed phase shift, we perform calculations where we once switch off the field-driven mixing to $\Phi_0$ after the probe step and once where we leave it on.
The probe pulse is here delta-like, i.e., $E_\text{probe}(t)\propto \delta(t-\tau)$.
We again reduce krypton to a two-level atom (excluding the $4s$ orbital) as described in Sec.~\ref{sec:res_residual}.
In this two-level system, $\phi_T^\text{ion}(\tau)=\phi_T^\text{residual}(\tau)=0$ such that the only phase shift that can occur is $\phi_T^\text{ground}(\tau)$.

\begin{figure}[ht!]
\begin{center}
  \rmpdfinfo
  \includegraphics[width=\linewidth]{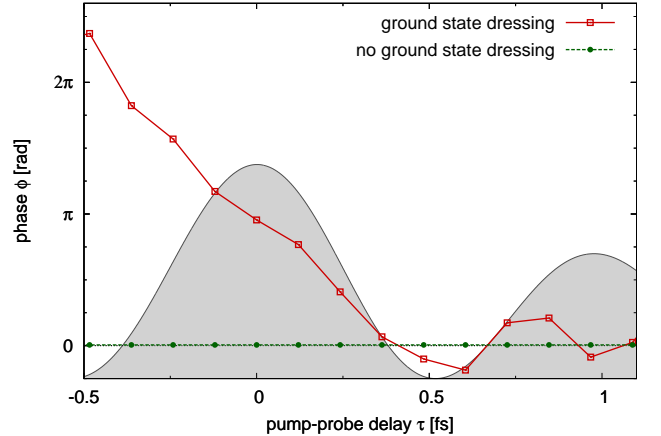}
  \caption{(color online) 
    The phase shifts $\phi_T(\tau)$ are shown for the transition $T=[4p^{1/2}_{3/2}]^{-1} \rightarrow [3d^{1/2}_{5/2}]^{-1}$. 
      The TDCIS calculations were done once with (red solid) and once without (green dashed) field-driving mixing with the neutral ground state $\Phi_0$ after the probe pulse.
      In both cases, no dressing of the ionic states can occur (see text for details).
  }
  \label{fig:phase_dressed-ground}
\end{center}
\end{figure}

In Fig.~\ref{fig:phase_dressed-ground}, the phase $\phi_{T}(\tau)$ is shown with and without the mixing to the neutral ground state.
If we set $\sbra{\Phi_0}\hat z\sket{\Phi^a_i}=0$ (i.e., $\phi_T^\text{ground}=0$) after the probe step, the phase shift totally disappears for all pump-probe delays.
When including the ground state mixing, we obtain again the usual behavior of $\phi_{T}(\tau)$.
Hence, we conclude that the main source of $\phi_T(\tau)$, which deforms the transition lines in the transient absorption spectrum (cf. Fig.~\ref{fig:tas_3d}), is the field-induced dressing of the entire $N$-electron system---particularly the mixing with the neutral ground state.
%

\section{\label{sec:conclusion} Conclusion}
We have described our theoretical model, namely a time-dependent configuration-interaction singles (TDCIS) approach, which has been implemented in the software package \textsc{xcid}.
We extended our method and included spin-orbit splitting for the occupied orbitals.
This extension leads to new symmetry classes and, therefore, new matrix elements.
With the help of this multi-electron approach, we are able to study attosecond transient absorption experiments for overlapping pump and probe pulses from first principles.
The pump pulse as well as the probe pulse are treated non-perturbatively.

Transient absorption spectroscopy with overlapping pulses makes it possible to study the tunnel ionization dynamics with a sub-cycle resolution.
We find that the hole populations extracted from the transient absorption spectrum are in close connection to the instantaneous hole populations even though this relations is strictly speaking only true for non-overlapping pump and probe pulses.

The strong deformations in the transient absorption lines, which appear during the ionization process, can be fully understood by phase shifts in the induced ionic dipole oscillations.
We find that the phase shift due to the dressing of the ion by the pump pulse is not sufficient to account for the observed line deformations.
We also excluded the residual Coulomb interaction between the ionized electron and the remaining ion as a possible source.

The main contribution to the phase shift comes from field-induced mixing of the excited $N$-electron states with the neutral ground state, which is highly non-perturbative.
This dressing mechanism creates a coupling between the ionized electron and the ion, which makes the ionic subsystem dependent on the state of the ionized electron and vice versa.
This dressing mechanism is quite peculiar, since the ionized electron was believed to be a spectator, since the probe pulse does only affect the ionic states. 
We are not aware that this effect has been discussed in the literature before.

The non-perturbative mixing to the neutral ground state affects also the phase relations between ionic states.
These phase relations are particularly important for the hole dynamics in an atom or molecule~\cite{lunnemann_ultrafast_2009}.
By varying the pump-probe delay, the phases between ionic states can be influenced, and transient absorption spectroscopy provides a way to ``read out'' these phases.

\acknowledgments
This work has been supported by the Deutsche Forschungsgemeinschaft (DFG) 
under grant No. SFB 925/A5.
We thank Jan Malte Slowik for helpful discussions.

\appendix

\section{Dipole Polarizability}
\label{app:polar}

The dipole polarizability of a system that is in the state $S$, can be obtained within perturbation theory and reads~\cite{MiSa-JPB-2010}
\begin{equation}
  \label{eq:polarizability}
  \alpha_{m,n} (S)
  = 
  2 \sum_I \frac{\bra{S} x_m \ket{I}\bra{I} x_n \ket{S}}{E_I-E_S}
  ,
\end{equation}
where $E_S$ and $E_I$ are the energies of the states $S$ and $I$, 
respectively, and $\sum_I$ stands for the sum over all intermediate states.
If $S$ is an eigenstate with magnetic quantum number $M_J$, 
the polarizability entries $\alpha_{x,x}(S)$ and $\alpha_{y,y}(S)$ are equal 
($\alpha_{x,x}(S) = \alpha_{y,y}(S)$)~\cite{BeRo-AdvChemPhys-1966}.
The formulas for $\alpha_{x,x}(S)$ and $\alpha_{z,z}(S)$ read 
\cite{BeRo-AdvChemPhys-1966}:
\begin{widetext}
\begin{subequations}
\label{eq:pol_symm}
\begin{align}
  \label{eq:pol_symm_xx}
  \alpha_{x,x}(S)
  =&\
  A_J\ [J(J-1)+M_J^2]
  &\hskip-4mm&+
  B_J\ [J(J+1)-M_J^2]
  &\hskip-4mm&+
  C_J\ [(J+1)(J+2)+M_J^2]
  ,
\\
  \label{eq:pol_symm_zz}
  \alpha_{z,z}(S)
  =&\
  2A_J\ [J^2-M_J^2]
  &\hskip-4mm&+
  2B_J\ M_J^2
  &\hskip-4mm&+
  2C_J\ [(J+1)^2 - M_J^2]
  ,
\end{align}
\end{subequations}
\end{widetext}
where the constants $A_J,B_J$,and $C_J$ stand for specific sums over intermediate
states $I$ with $J_I=J_S-1,J_I=J_S,$ and $J_I=J_S+1$, respectively.

For singly ionized, atomic krypton, we are interested in the polarizabilities for ionic states with a hole in the $4p$ or the $3d$ orbital manifolds.
Using {\sc DALTON}, we calculated polarizabilities that do not include spin-orbit coupling.
In order to obtain polarizabilities for the spin-orbit-coupled ionic states
$[4p^{m_j}_{j}]^{-1}$ and $[3d^{m_j}_{j}]^{-1}$ we perform first order 
perturbation theory for degenerate states.
The diagonal polarizability entries of the spin-orbit-coupled ionic states $[NL^{M_J}_J]^{-1}$ expressed with the polarizabilities of the non-relativistic states $[NL_{M_L}]^{-1}$ read
\begin{align}
  \label{eq:pol_LS-trafo}
  \alpha_{n,n}\left(\Big.[NL^{M_J}_J]^{-1}\right)
  =&
  \sum_{\sigma,M_L} \
  [C^{J,M_J}_{L,M_L;1/2,\sigma}]^2
\\\nonumber 
  &\times
  \alpha_{n,n}\left(\Big.[NL_{M_L}]^{-1}\right)
  .
\end{align}
With the help of Eqs.~\eqref{eq:pol_symm} and \eqref{eq:pol_LS-trafo} only two
polarizability entries $\alpha_{n,n}(S), n\in\{x,y,z\}$ from states with the same $N$ and $L$ quantum numbers are needed and all other diagonal polarizability 
entries of any state $S'$ can be obtained as long as $S'$ has the same $N$ and $L$ quantum numbers.
\begin{table}
  \caption{\label{tab:polar_dalton} Static dipole polarizabilities $\alpha_{x,x}$ and 
    $\alpha_{z,z}$ are shown for two Kr$^+$ states. 
    All other polarizabilities can be deduced from these values.
    Polarizabilities obtained by the CIS and CASSCF methods are compared.
    All values are given in atomic units with a precision up to the second digit.
  }  
  \begin{ruledtabular}
  \begin{tabular}{ l | c c | c  c }
    & \multicolumn{2}{c|}{CIS} & \multicolumn{2}{c}{CASSCF}  \\
    & $\alpha_{x,x}$ & $\alpha_{z,z}$  &  $\alpha_{x,x}$ & $\alpha_{z,z}$ \\
    \hline
    $[4p_0]^{-1}$   &  0.01  &  3.13 & 10.77 & 10.46   \\  
    $[3d_0]^{-1}$   &  0.00  &  0.00 &  9.66 &  9.52   
  \end{tabular}
  \end{ruledtabular}
\end{table}
In Table~\ref{tab:polar_dalton} the calculated polarizabilities are shown that are 
used to obtain all polarizabilities in Table~\ref{tab:polar}.
For the CASSCF and CIS calculations we obtained converged results with 
the augmented correlation-consistent quintuple-zeta basis set including polarization functions (aug-cc-pV5Z)~\cite{WiWo-JCP-1999}.


\bibliographystyle{./apsrev4-1}
\bibliography{amo,books,solidstate,polar_refs}

\end{document}